\documentclass[12pt]{article}
\usepackage[margin=0.7in]{geometry}
\usepackage{graphicx} 
\usepackage{authblk} 
\usepackage{setspace}
\usepackage{natbib}
\usepackage{float}
\usepackage{placeins}
\usepackage{amsthm,amsmath,amssymb,amscd}
\usepackage{booktabs}
\usepackage{threeparttable}
\usepackage{xcolor}
\DeclareMathOperator{\logit}{logit}
\DeclareMathOperator{\expit}{expit}
\newcommand{\mploco}{MPLOCO}
\newtheorem{theorem}{Theorem}

\title{Bridging extrinsic and intrinsic variable importance}
\author[1,*]{Yucheng Zhao}
\author[2,3,4]{Brian~D. Williamson}
\affil[1]{Department of Statistics, University of Washington}
\affil[2]{Biostatistics Division, Kaiser Permanente Washington Health Research Institute}
\affil[3]{Vaccine and Infectious Disease Division, Fred Hutchinson Cancer Center}
\affil[4]{Department of Biostatistics, University of Washington}
\affil[*]{Corresponding Author: Department of Biostatistics and Informatics, University of Colorado Anschutz, Aurora, CO 80045. Email: yucheng.zhao@ucdenver.edu}
\date{\today}

\begin{document}

\maketitle

\begin{abstract}
    Variable importance may describe either intrinsic predictive information in a population or extrinsic importance for a fitted prediction rule. Quantifying the uncertainty in variable importance estimates is critical for interpretation. Methods for estimating intrinsic variable importance (we will refer to these as VIMP) and the minipatch leave-one-covariate-out procedure (MPLOCO) target intrinsic and extrinsic importance, respectively, and provide methods for computing standard errors. These two approaches have a shared structure, comparing prediction performance with and without features, but the relationship between them has not been formally characterized. We establish conditions under which the two perspectives align. Under squared-error loss, if the fitted full and reduced learners converge to their oracle counterparts sufficiently fast, then MPLOCO is asymptotically equivalent to VIMP. We provide further conditions extending this result to general loss functions and formalize grouped MPLOCO for potentially overlapping feature groups. Through simulations, we show that VIMP and MPLOCO agree most closely when the fitted learner is well aligned with the data-generating mechanism. In a high-dimensional grouped simulation, both procedures identified the signal-containing groups. In an analysis of HIV-1 VRC01 neutralization sensitivity, both methods placed the same three biologically relevant feature groups among their highest-ranked groups. These results clarify when intrinsic and extrinsic importance can be interpreted similarly and when they provide complementary information.
\end{abstract}

\doublespacing
\section{Introduction}

Modern prediction algorithms are increasingly used in biomedical, clinical, and social science applications, where investigators often seek both accurate prediction and scientific insight into the variables driving prediction \citep{lundberg2018explainable,magaret2019prediction,guess2023social,molnar2020interpretable}. For example, studies of HIV-1 neutralization sensitivity seek both to predict viral susceptibility and to identify sequence features associated with neutralization response and prevention efficacy \citep{magaret2019prediction,juraska2024prevention}. There has been a large increase in methods for estimating variable importance in recent years.

The term \textit{variable importance} encompasses two distinct targets, which we will call \textit{extrinsic} and \textit{intrinsic} importance. Extrinsic importance describes the importance of a variable in a fitted prediction model. Examples include regression coefficients, changes in $R^2$, likelihood comparisons, or ANOVA decompositions in generalized linear regression models \citep{johnson2004history,gromping2009variable}; coefficients in penalized regression models \citep{tibshirani1996regression}; impurity and permutation importance for random forests \citep{breiman2001random,strobl2007bias}; and connection-weight methods for neural networks \citep{olden2004accurate}. In recent work, extrinsic measures that apply to broad classes of algorithms have been developed, including model reliance and permutation importance \citep{fisher2019all}, leave-one-covariate-out importance \citep{lei2018distribution}, ANOVA decompositions \citep{owen2013variance}, and Shapley-based measures including SHAP \citep{lundberg2017unified}. Intrinsic importance describes the importance of a variable in the population that gave rise to the data, and is not tied to a particular fitted algorithm. Examples include differences in the average treatment effect \citep{van2005statistical,hubbard2013time,diaz2015variable}, nonparametric $R^2$ \citep{williamson2021nonparametric}, and a class of measures that includes the above measures and classification accuracy and area under the receiver operating characteristic curve \citep{williamson2023general}, with extensions to Shapley values \citep{williamson2020efficient}, survival outcomes \citep{wolock2025assessing}, and treatment effect heterogeneity \citep{hines2022variable}. Extrinsic and intrinsic importance provide complementary information, and often both are of interest in a data analysis.

Because variable importance is estimated from data, uncertainty quantification is essential for interpretation. Point estimates alone may make variables appear meaningfully different in importance when the observed differences are largely attributable to sampling variability, especially in modest samples or when predictive information is distributed across correlated features. Many of the extrinsic approaches described above do not yield valid confidence intervals. This concern has motivated inferential procedures and the construction of confidence intervals for variable importance. The intrinsic procedures above support valid inference based on semiparametric theory \citep{williamson2023general}. We will refer to these approaches as VIMP approaches. A recent approach for creating confidence intervals for extrinsic importance \citep{gan2022model}, which we will call MPLOCO, uses random subsampling of observations and features (minipatches) to construct point estimates and confidence intervals for extrinsic variable importance. 

Intrinsic VIMP and extrinsic MPLOCO are complementary approaches to estimating and making inference on importance. VIMP asks how much population-level oracle predictiveness is lost when a feature or group is unavailable. MPLOCO asks how much the same exclusion changes predictive performance for a fitted minipatch learner and, before taking an asymptotic limit, depends on the observed training data, base learner, and minipatch construction. Both questions can be scientifically relevant: investigators may care about the intrinsic predictive information carried by a variable while also wanting to understand how the prediction procedure actually deployed uses that information. Moreover, both approaches compare predictive performance with and without a feature under a chosen loss. This shared structure suggests that MPLOCO may approximate VIMP when the fitted full and reduced prediction functions are sufficiently close to their oracle counterparts.

In this paper, we formalize this connection, bridging the gap between extrinsic and intrinsic importance. We introduce conditions under which the MPLOCO estimand converges to the corresponding population-level VIMP estimand and characterize when their point estimators and variance estimators have the same asymptotic behavior. We then investigate the relationship empirically in simulations, including settings in which the learner is well specified or misspecified and a high-dimensional setting with overlapping feature groups. Finally, we compare the two procedures in an analysis of genetic features associated with VRC01 neutralization sensitivity. Our aim is to clarify when the procedures yield similar rankings, point estimates, and uncertainty quantification, and when their differences should instead be interpreted as complementary information about population-level predictive value and the behavior of a fitted learner.

The remainder of the paper is organized as follows. We define the two estimands in Section 2 and develop their theoretical connection. In Section 3, we present results from numerical experiments, investigate variable importance in an HIV-1 dataset in Section 4 and provide conclusions in Section 5.

\section{Methods}
\subsection{Setup and notation}

We use similar notation to \citet{williamson2023general}. Let $Z_i=(X_i, Y_i), i=1,\ldots,n$ denote independent observations drawn from an unknown data-generating distribution $P_0$. The covariate vector is $X_i=(X_{i1},\ldots,X_{ip})^\top\in\mathcal{X}$, and $Y_i$ denotes the outcome. Let $s\subseteq\{1,\ldots,p\}$ denote a target feature or a set of target features, and let $X_{-s}$ denote the covariate vector after excluding the features indexed by $s$.

For a prediction function $f$ and loss function $\ell$, define the population expected loss \[ R_0(f) = E_0\left[\ell\{Y,f(X)\}\right].\] Let $\mathcal{F}$ denote a sufficiently rich class of prediction functions based on the full covariate vector, and let $\mathcal{F}_{-s}\subseteq\mathcal{F}$ denote the class of prediction functions whose evaluation depends only on $X_{-s}$. The corresponding oracle prediction functions are
\[f_0 \in \arg\min_{f\in\mathcal{F}}R_0(f)\] and \[f_{0,-s} \in \arg\min_{f\in\mathcal{F}_{-s}}R_0(f).\]
Thus, $f_0$ is the optimal population prediction function using all features, while $f_{0,-s}$ is the optimal population prediction function after excluding the target features.

Our primary theoretical development uses squared-error loss, $\ell\{y,f(x)\}=\{y-f(x)\}^2,$ for ease of exposition. Our results can be generalized to other loss functions, as we show below. Under squared-error loss and a sufficiently rich function class, $f_0(x)=E_0(Y\mid X=x)$ and $f_{0,-s}(x_{-s}) = E_0(Y\mid X_{-s}=x_{-s}).$ In many other cases the conditional mean outcome given covariates (or a simple function of this conditional mean) remains the optimal prediction function, including classification accuracy, cross-entropy (deviance), and the area under the receiver operating characteristic curve \citep{williamson2023general}. 

\subsection{Intrinsic population-level VIMP}

Following \citet{williamson2023general}, intrinsic variable importance is defined as the loss in oracle predictiveness caused by excluding the target features. On the risk scale, the intrinsic population-level importance of $X_s$ is \[\theta_{0,1} = R_0(f_{0,-s})-R_0(f_0).\] A positive value indicates that the optimal prediction function based only on $X_{-s}$ has larger expected loss than the optimal prediction function based on all features; by construction, $\theta_{0,1} \geq 0$. Because $\theta_{0,1}$ is a function only of the optimal prediction functions and $P_0$, it is a property of the population distribution $P_0$ rather than of a particular fitted algorithm.

In practice, $f_0$ and $f_{0,-s}$ are unknown and are estimated using full and reduced prediction algorithms. Let $\widehat f_n$ denote an estimator of $f_0$, and let $\widehat f_{n,-s}$ denote an estimator of $f_{0,-s}$. To facilitate inference under the zero-importance null hypothesis, VIMP may be implemented using sample splitting, where $R(f_0)$ and $R(f_{0,-s})$ are estimated on independent halves of the data. If flexible machine learning models are used to estimate $f_0$ and $f_{0,-s}$, cross-fitting removes the need for possibly restrictive assumptions on these models \citep{williamson2023general}.

Let $v(i)$ denote the validation fold containing observation $i$, and let $\widehat f_n^{(-v(i))}$ and $\widehat f_{n,-s}^{(-v(i))}$ denote the full and reduced learners trained without fold $v(i)$. Under squared-error loss, define the cross-fitted observation-level contrast
\[\widehat g_{i,1} = \left\{ Y_i- \widehat f_{n,-s}^{(-v(i))}(X_{i,-s}) \right\}^2 - \left\{ Y_i- \widehat f_n^{(-v(i))}(X_i) \right\}^2.\]
The corresponding VIMP estimator may be written as \[\widehat\theta_{n,1} = \frac{1}{n} \sum_{i=1}^n \widehat g_{i,1},\] with its standard error obtained from the influence-function-based procedure described by \citet{williamson2023general}.

For later comparison, define the oracle loss contrast \[g_0(Z) = \left\{ Y-f_{0,-s}(X_{-s}) \right\}^2 - \left\{ Y-f_0(X) \right\}^2.\] Then \[E_0\{g_0(Z)\}=\theta_{0,1}.\]
Under conditions requiring local quadratic optimality and differentiability of the risk functional, sufficiently fast convergence to both full and reduced prediction-function estimators to their corresponding oracle functions at rates faster than $n^{-1/4}$, and control of the empirical-process remainder terms, \citet{williamson2023general} establish that
\[\sqrt{n} \left( \widehat\theta_{n,1}-\theta_{0,1} \right) \overset{d}{\longrightarrow} N(0,\sigma_1^2),\]
where $\sigma_1^2 = \operatorname{Var}_0\{g_0(Z)\}.$ This result provides inference for the intrinsic population-level importance parameter.

\subsection{Extrinsic fitted-function MPLOCO}
\label{sec:mploco}

\mploco{} \citep{gan2022model} defines importance relative to a fitted minipatch ensemble constructed from the observed sample $\mathcal{D}_n=\{Z_i:i=1,\ldots,n\}$. Unlike the sample-splitting implementation of VIMP described above, which estimates full and reduced predictiveness using separate subsets of observations, MPLOCO constructs a minipatch ensemble from the full observed sample. The minipatches are random subsets of both observations and features; the algorithm of interest is fitted to each minipatch. For each observation, predictions are obtained only from minipatches that exclude that observation. The full prediction is aggregated over all such minipatches, whereas the reduced prediction is obtained from the subset of these minipatches that additionally exclude the target feature(s).

A minipatch consists of a subset of both observations and a subset of features. For each minipatch $b=1,\ldots,B$, let $I_b\subseteq\{1,\ldots,n\}$ denote the sampled observations and let $F_b\subseteq\{1,\ldots,p\}$ denote the sampled features. A specified base prediction algorithm, such as lasso \citep{tibshirani1996regression} or random forest \citep{breiman2001random}, is fit using the observations indexed by $I_b$ and features indexed by $F_b$, producing a minipatch learner $\widehat f_b$.

For observation $i$, the full prediction is obtained by averaging predictions from minipatches that exclude observation $i$:
\[\widehat f_{-i}(X_i) = \frac{ \sum_{b=1}^B \mathbf{1}(i\notin I_b) \widehat f_b(X_i) }{ \sum_{b=1}^B \mathbf{1}(i\notin I_b) }.\]
To assess the importance of the target feature set $s$, the leave-$s$-out prediction averages only over minipatches whose feature sets contain none of the target features:
\[\widehat f_{-i,-s}(X_i) = \frac{ \sum_{b=1}^B \mathbf{1}(i\notin I_b) \mathbf{1}(F_b\cap s=\emptyset) \widehat f_b(X_i) }{ \sum_{b=1}^B \mathbf{1}(i\notin I_b) \mathbf{1}(F_b\cap s=\emptyset) }.\]

The observation-level \mploco{} contrast is \[\widehat\Delta_{i,s} = \ell\{Y_i,\widehat f_{-i,-s}(X_i)\} - \ell\{Y_i,\widehat f_{-i}(X_i)\},\] and the \mploco{} estimator is \[\widehat\theta_{n,2} = \frac{1}{n} \sum_{i=1}^n \widehat\Delta_{i,s}.\] Its empirical variance estimator is \[\widehat\sigma_{2,n}^2 = \frac{1}{n-1} \sum_{i=1}^n \left( \widehat\Delta_{i,s} - \widehat\theta_{n,2} \right)^2,\] leading to the Wald-type confidence interval \[\widehat\theta_{n,2} \pm z_{1-\alpha/2} \frac{\widehat\sigma_{2,n}}{\sqrt n}.\]

Unlike $\theta_{0,1}$, the \mploco{} target is defined relative to the fitted ensemble. Let $f_n$ denote the full fitted minipatch prediction function and let $f_{n,-s}$ denote its leave-$s$-out counterpart. The conditional fitted-function target is \[\theta_{0,2}(f_n) = E_0 \left[ \ell\{Y,f_{n,-s}(X_{-s})\} - \ell\{Y,f_n(X)\} \mid \mathcal{D}_n \right].\] This target depends on the observed training data, the base learner, the observation and feature minipatch sizes, and other tuning choices used to construct the ensemble.

\citet{gan2022model} establish asymptotic normality of $\widehat\theta_{n,2}$ under conditions that ensure stability of the minipatch procedure. Specifically, the loss function is assumed to be Lipschitz in its prediction argument; predictions from models fit on different minipatches are required to have uniformly bounded differences; the observation and feature minipatch sizes must satisfy rate conditions that prevent individual minipatches from having excessive influence; and the number of sampled minipatches must grow sufficiently quickly for the additional randomness from minipatch sampling to be asymptotically negligible. Their result also requires uniform integrability of the standardized observation-level importance quantities. Under these conditions,
\[\frac{ \sqrt n \left\{ \widehat\theta_{n,2} - \theta_{0,2}(f_n) \right\} }{ \sigma_{2,n} } \overset{d}{\longrightarrow} N(0,1),\] and \[\frac{ \widehat\sigma_{2,n}^2 }{ \sigma_{2,n}^2 } \overset{P}{\longrightarrow} 1.\]
Thus, under these conditions, MPLOCO provides asymptotically valid inference for its fitted-function importance target $\theta_{0,2}(f_n)$.

\subsection{Connecting intrinsic VIMP and fitted-function MPLOCO}

The intrinsic VIMP and fitted-function \mploco{} targets are different quantities. Nevertheless, they have the same full-versus-reduced risk structure, and both rely on prediction-function estimators that may be similar in practice. This shared structure makes it possible to describe conditions under which the fitted-function target approaches the intrinsic population-level target. In this section, $f_n$ and $f_{n,-s}$ denote estimators of $f_0$ and $f_{0,-s}$, respectively, which may be obtained using minipatch ensembles or another method such as cross-fitting.

For simplicity, we first present the connection under squared-error loss. This loss yields an exact excess-risk identity and satisfies the smooth, locally quadratic risk conditions used for VIMP inference. The corresponding extension to more general losses is stated after Theorem~\ref{thm:vimp_mploco_alignment}, with additional details in the Supplement. Under squared-error loss, define \[g_n(Z)=\left\{Y-f_{n,-s}(X_{-s})\right\}^2-\left\{Y-f_n(X)\right\}^2.\] Conditional on the fitted learners, \[E_0\{g_n(Z)\mid\mathcal{D}_n\}=\theta_{0,2}(f_n).\] Recall that \[g_0(Z)=\left\{Y-f_{0,-s}(X_{-s})\right\}^2-\left\{Y-f_0(X)\right\}^2,\] so that $E_0\{g_0(Z)\}=\theta_{0,1}$. For the fitted-contrast formulation used below, write \[\sigma_{2,n}^2=\operatorname{Var}_0\{g_n(Z)\mid\mathcal D_n\}.\]

\paragraph{Bridge assumptions.}

We use the following conditions to connect VIMP and MPLOCO.

\begin{enumerate}
\item[(C1)] \textbf{Convergence of the full and reduced learners.} The fitted full and reduced prediction functions satisfy $\|f_n-f_0\|_{L_2(P_0)} = o_P(n^{-1/4})$ and $\|f_{n,-s}-f_{0,-s}\|_{L_2(P_0)} = o_P(n^{-1/4}).$

\item[(C2)] \textbf{Convergence of the loss contrasts.} The fitted and oracle observation-level loss contrasts satisfy $\|g_n-g_0\|_{L_2(P_0)} \overset{P}{\longrightarrow} 0.$

\item[(C3)] \textbf{Valid estimator-specific inference.} The VIMP estimator is asymptotically normal around $\theta_{0,1}$ and the \mploco{} estimator is asymptotically normal around $\theta_{0,2}(f_n)$, and $\widehat\sigma_{2,n}^2$ consistently estimates its asymptotic variance $\sigma_{2,n}^2$.

\item[(C4)] \textbf{Nondegenerate limiting variance.} The oracle loss contrast satisfies $0 < \sigma_1^2 = \operatorname{Var}_0\{g_0(Z)\} < \infty.$
\end{enumerate}

Condition (C1) requires both fitted learners to approximate their corresponding oracle prediction functions. Condition (C2) is stated separately from (C1) because $L_2(P_0)$ convergence of prediction functions does not by itself guarantee $L_2(P_0)$ convergence of squared-loss contrasts when outcomes or predictions are unbounded. Under squared-error loss, (C2) follows from (C1) if the outcome and the fitted and oracle predictions are uniformly bounded. It also follows under suitable fourth-moment bounds together with $L_4(P_0)$ convergence of the full and reduced learners. These sufficient conditions are stated and verified in the Supplement.

Condition (C3) combines the source inference results for the two estimators. For VIMP, the predictiveness functional must satisfy the local quadratic optimality and differentiability conditions of \citet{williamson2023general}, and the nuisance-estimation remainder must be controlled. Squared-error risk satisfies the required exact quadratic expansion; mean absolute error is not used here because its risk is generally nonsmooth and does not automatically satisfy these conditions. For \mploco{}, the source assumptions of \citet{gan2022model} include a loss that is Lipschitz in its prediction argument, bounded differences between predictions from different minipatches, rate restrictions on the observation and feature minipatch sizes, sufficiently many sampled minipatches, and a uniform-integrability condition. Squared-error loss satisfies the Lipschitz requirement on a bounded outcome and prediction range, but not automatically on an unbounded range. The full estimator-specific conditions are recorded in the Supplement. Condition (C4) restricts attention to the nondegenerate setting in which ordinary Wald inference is applicable.

\paragraph{Alignment under squared-error loss.}

Because $f_0(x)=E_0(Y\mid X=x)$ and
$f_{0,-s}(x_{-s})=E_0(Y\mid X_{-s}=x_{-s})$, and because $R_0$ evaluates risk on an independent test observation, the following identities hold exactly conditional on the fitted learners:
\[R_0(f_n)-R_0(f_0)=\|f_n-f_0\|_{L_2(P_0)}^2\]and\[R_0(f_{n,-s})-R_0(f_{0,-s})=\|f_{n,-s}-f_{0,-s}\|_{L_2(P_0)}^2.\]
Thus, no $o_P(n^{-1/2})$ remainder is part of either identity. Consequently,
\[\begin{aligned}\theta_{0,2}(f_n)-\theta_{0,1}={}&\left[R_0(f_{n,-s})-R_0(f_{0,-s})\right]\\&-\left[R_0(f_n)-R_0(f_0)\right]\\={}&\|f_{n,-s}-f_{0,-s}\|_{L_2(P_0)}^2-\|f_n-f_0\|_{L_2(P_0)}^2.\end{aligned}
\]

The next theorem describes the alignment of VIMP and MPLOCO under Conditions (C1)--(C4) and the additional conditions required for valid inference on the VIMP and MPLOCO estimands.

\begin{theorem}[Alignment of intrinsic VIMP and \mploco{}]
\label{thm:vimp_mploco_alignment}
Suppose squared-error loss is used and Conditions (C1)--(C4) hold. Then:
\begin{enumerate}
\item[(i)] $\theta_{0,2}(f_n) - \theta_{0,1} \overset{P}{\longrightarrow} 0;$

\item[(ii)] $\sqrt n \left\{ \theta_{0,2}(f_n) - \theta_{0,1} \right\} = o_P(1);$

\item[(iii)] $\sigma_{2,n}^2 \overset{P}{\longrightarrow} \sigma_1^2$ and  $\widehat\sigma_{2,n}^2 \overset{P}{\longrightarrow} \sigma_1^2;$

\item[(iv)] $\sqrt n \left( \widehat\theta_{n,2} - \theta_{0,1} \right) \overset{d}{\longrightarrow} N(0,\sigma_1^2).$
\end{enumerate}
\end{theorem}
Therefore, when the fitted full and reduced learners approach their oracle counterparts sufficiently quickly, the VIMP and \mploco{} estimators have the same first-order limiting distribution, despite being constructed using different inferential procedures.

The theorem does not imply that the two procedures must agree in finite samples. Differences may remain when the fitted learners do not approximate the oracle functions well, when the full and reduced learners converge at different rates, or when the minipatch construction changes the prediction functions being evaluated. These differences are particularly relevant in modest-sample, high-dimensional settings.

\paragraph{Extension beyond squared-error loss.}

The squared-error result uses the exact excess-risk identity above. For a more general loss, the same alignment conclusion holds if
\[\sqrt n \left[ \left\{ R_0(f_{n,-s})-R_0(f_{0,-s}) \right\} - \left\{ R_0(f_n)-R_0(f_0) \right\} \right] = o_P(1)\]
and if the corresponding fitted and oracle observation-level loss contrasts converge in $L_2(P_0)$. These conditions replace the special quadratic identity used for mean squared error. A general-loss version of the result and its proof are provided in the Supplement.

\subsection{Grouped variable importance}\label{sec:group_vim}

While the preceding definitions apply to either individual features or groups of features, computing variable importance for feature groups requires special consideration. Let $\mathcal{G} = \{G_1,\ldots,G_K\}$
denote a collection of feature groups, where $G_k\subseteq\{1,\ldots,p\}$. We allow the groups to overlap, so a single feature may belong to more than one group.

For target group $G_k$, the intrinsic grouped VIMP parameter is \[\theta_{0,1}^{(k)} = R_0(f_{0,-G_k})-R_0(f_0),\] where $f_{0,-G_k}$ is the oracle prediction function based on all features except those belonging to $G_k$. In practice, grouped VIMP is estimated by fitting a full learner using all features and a reduced learner after removing all features in the target group. Prediction risk is then evaluated using sample splitting or cross-fitting, with standard errors obtained from the VIMP inference procedure. This is identical to the approach used for individual features.

The minipatch construction naturally extends to grouped feature importance. Instead of sampling individual features, we sample groups of features and include all variables belonging to the sampled groups in each minipatch. For minipatch $b$, let $I_b$ denote the sampled observations and let $S_b\subseteq\{1,\ldots,K\}$ denote the sampled groups. The feature set used in the minipatch is $F_b = \bigcup_{k\in S_b}G_k.$
When the groups overlap, duplicated features are included only once. A base learner is then fit using observations $I_b$ and features $F_b$. For each observation $i$, the full out-of-minipatch prediction averages over minipatches satisfying $i\notin I_b$. For target group $G_k$, the leave-group-out prediction is
\[\widehat f_{-i,-G_k}(X_i) = \frac{ \sum_{b=1}^B \mathbf{1}(i\notin I_b) \mathbf{1}(F_b\cap G_k=\emptyset) \widehat f_b(X_i) }{ \sum_{b=1}^B \mathbf{1}(i\notin I_b) \mathbf{1}(F_b\cap G_k=\emptyset) }.\]
The grouped construction differs from the individual-feature version only in the exclusion criterion. For an individual feature, a minipatch is excluded when the target feature is included in $F_b$. For a group $G_k$, a minipatch contributes to the reduced prediction only when none of the variables in the group are included, i.e., $F_b\cap G_k=\emptyset$. When groups overlap, a variable shared by multiple groups is treated as belonging to each group. Therefore, a minipatch containing a shared variable is excluded from the reduced prediction for any target group containing that variable.

The grouped observation-level contrast is \[\widehat\Delta_{i,G_k} = \ell\{Y_i,\widehat f_{-i,-G_k}(X_i)\} - \ell\{Y_i,\widehat f_{-i}(X_i)\},\] and grouped \mploco{} is estimated by \[\widehat\theta_{n,2}^{(k)} = \frac{1}{n} \sum_{i=1}^n \widehat\Delta_{i,G_k}.\] A Wald-type confidence interval is constructed from the empirical standard deviation of the observation-level contrasts.

The corresponding fitted-function target is \[\theta_{0,2}^{(k)}(f_n) = R_0(f_{n,-G_k})-R_0(f_n).\]
Both grouped VIMP and grouped \mploco{} therefore assess the predictive contrast obtained when all features in the target group are made unavailable. VIMP constructs this contrast by refitting a reduced learner, whereas grouped \mploco{} constructs it by selecting the subset of already fitted minipatch learners whose feature sets contain none of the target-group features. The alignment result in the previous subsection applies with $s=G_k$ whenever the corresponding full and reduced convergence conditions hold.

\section{Numerical experiments}
\subsection{Investigating the importance of individual features}

The first numerical experiment focused on individual-feature importance. The goal was to compare intrinsic VIMP and \mploco{} when the fitted learner was either well aligned or poorly aligned with the data-generating mechanism. Throughout this experiment, variable importance was measured using differences in mean squared error, matching the theoretical development in Section~2 and placing the two procedures on the same loss scale. Positive values therefore indicate that removing the feature worsens predictive performance.

For each of 200 Monte Carlo replications, we generated $n = 1000$ independent observations with $p=60$ features. The feature vector was generated as $X_i=(X_{i1},\ldots,X_{ip})^\top,$ $X_{ij}\overset{\mathrm{iid}}{\sim}N(0,1).$ The outcome was generated from a binary regression model, $Y_i\mid X_i\sim
\operatorname{Bernoulli}(\mu_i)$, where $\mu_i=\expit(2X_{i1}-2X_{i2}).$ Thus, only $X_1$ and $X_2$ were predictive of the outcome, while $X_3, \ldots, X_{60}$ were noise variables. Because $X_1$ and $X_2$ had coefficients of equal magnitude and the covariates were independently and symmetrically distributed, the two signal variables had the same intrinsic importance. We therefore report results for $X_1$ as the representative signal feature.

For MPLOCO, each ensemble contained $B=500$ minipatches. Each minipatch sampled $n_{\mathrm{ratio}}=0.8$ of the observations and $m_{\mathrm{ratio}}=0.8$ of the features. We considered a logistic generalized linear model (GLM), a linear model (LM), and three prespecified random-forest configurations denoted strong, middle, and weak. The three random-forest configurations differed in tree depth and ensemble size. The strong, middle, and weak random forests used $(ntree, nodesize)$ settings of $(200,20)$, $(100,100)$, and $(50,300)$, respectively. When both signal features were available, the logistic GLM was correctly specified for the full conditional mean, whereas the LM provided a parametric misspecification benchmark because it imposed an identity-link linear mean on a Bernoulli response. The random-forest configurations represented different levels of predictive strength. For VIMP, we used the logistic GLM, LM, and a default random forest, with two-fold sample splitting. The VIMP estimates were reported on the same unscaled mean-squared-error difference scale as MPLOCO. All procedures used two-sided Wald-type confidence intervals with nominal coverage 95\%.


The intrinsic population target was evaluated numerically as $\theta_{0,1}=0.070967.$ For each learner $a$, we also calculated a learner-specific Monte Carlo reference value, denoted by $\theta_{0,2}^{(a)}$, for the corresponding fitted-function importance. These reference values were approximated using large independent Monte Carlo test samples of size 60,000 for the parametric learners and 10,000 for the random-forest learners. For target $k\in\{1,2\}$, empirical bias and mean squared error were calculated as 
\[\widehat{\operatorname{Bias}}_k=\frac{1}{R}\sum_{r=1}^{R}\widehat\theta^{(r)}-\theta_{0,k}^{(a)},\qquad\widehat{\operatorname{MSE}}_k=\frac{1}{R}\sum_{r=1}^{R}\left\{\widehat\theta^{(r)}-\theta_{0,k}^{(a)}\right\}^{2},\qquad\theta_{0,1}^{(a)}\equiv\theta_{0,1}.\]
The empirical coverage probability was the proportion of the 200 confidence intervals containing the relevant reference value. As clarified in the table notes, coverage relative to the fixed learner-level value $\theta_{0,2}^{(a)}$ is descriptive rather than replication-specific conditional coverage.

\begin{table}[t]
\centering
\caption{Individual-feature importance results for the binary-response simulation with $R=200$, $n=1000$, $p=60$, $B=500$, and $(n_{\mathrm{ratio}},m_{\mathrm{ratio}})=(0.8,0.8)$. Panel A evaluates each procedure relative to the intrinsic target $\theta_{0,1}=0.070967$. Panel B evaluates the same estimates relative to the learner-specific fitted-function reference $\theta_{0,2}^{(a)}$.}
\label{tab:individual-binary}
\begin{threeparttable}
\small
\renewcommand{\arraystretch}{1.08}

\textit{Panel A: Performance relative to the intrinsic target
$\theta_{0,1}$}\par\smallskip
\begin{tabular}{llrrrrr}
\toprule
Method & Base learner & Mean estimate & Mean SE & Bias$_1$ & MSE$_1$ & Coverage$_1$ \\
\midrule
MPLOCO & Logistic GLM & 0.069198 & 0.005355 & $-0.001769$ & $4.078\times10^{-5}$ & 0.920 \\
MPLOCO & LM & 0.058780 & 0.007857 & $-0.012187$ & $2.464\times10^{-4}$ & 0.585 \\
MPLOCO & RF--strong & 0.054043 & 0.003903 & $-0.016924$ & $3.210\times10^{-4}$ & 0.080 \\
MPLOCO & RF--middle & 0.049595 & 0.003665 & $-0.021372$ & $4.852\times10^{-4}$ & 0.005 \\
MPLOCO & RF--weak & 0.036273 & 0.002961 & $-0.034695$ & $1.224\times10^{-3}$ & 0.000 \\
\addlinespace
VIMP & Logistic GLM & 0.071988 & 0.011407 & 0.001021 & $2.381\times10^{-4}$ & 0.880 \\
VIMP & LM & 0.061911 & 0.024068 & $-0.009056$ & $8.161\times10^{-4}$ & 0.900 \\
VIMP & RF & 0.059216 & 0.006906 & $-0.011751$ & $2.097\times10^{-4}$ & 0.575 \\
\bottomrule
\end{tabular}

\medskip
\textit{Panel B: Performance relative to the fitted-function reference
$\theta_{0,2}^{(a)}$}\par\smallskip
\begin{tabular}{llrrrr}
\toprule
Method & Base learner & $\theta_{0,2}^{(a)}$ & Bias$_2$ & MSE$_2$ & Coverage$_2$ \\
\midrule
MPLOCO & Logistic GLM & 0.070835 & $-0.001637$ & $4.033\times10^{-5}$ & 0.915 \\
MPLOCO & LM & 0.057063 & 0.001717 & $1.008\times10^{-4}$ & 0.865 \\
MPLOCO & RF--strong & 0.069048 & $-0.015005$ & $2.597\times10^{-4}$ & 0.110 \\
MPLOCO & RF--middle & 0.067189 & $-0.017594$ & $3.380\times10^{-4}$ & 0.035 \\
MPLOCO & RF--weak & 0.061210 & $-0.024938$ & $6.422\times10^{-4}$ & 0.000 \\
\addlinespace
VIMP & Logistic GLM & 0.070835 & 0.001153 & $2.384\times10^{-4}$ & 0.875 \\
VIMP & LM & 0.057063 & 0.004849 & $7.576\times10^{-4}$ & 0.910 \\
VIMP & RF & 0.069048 & $-0.009832$ & $1.683\times10^{-4}$ & 0.685 \\
\bottomrule
\end{tabular}

\begin{tablenotes}[flushleft]
\footnotesize
\item \textit{Notes.} Mean estimate and mean SE are Monte Carlo averages over the 200 replications. Bias$_k$ and MSE$_k$ are computed relative to target $k$, and Coverage$_k$ is the empirical coverage of the corresponding two-sided $95\%$ Wald interval. For the VIMP RF row, the reported $\theta_{0,2}^{(a)}$ is the RF--strong reference value used in the comparison.
Coverage$_2$ is descriptive because a fixed learner-level reference is used in place of the replication-specific conditional target $\theta_{0,2}(f_n)$.
\end{tablenotes}
\end{threeparttable}
\end{table}

Table~\ref{tab:individual-binary} shows the clearest agreement between the intrinsic and fitted-function perspectives under the correctly specified logistic GLM. In this setting, the intrinsic target $\theta_{0,1}=0.070967$ and the fitted-function reference $\theta_{0,2}^{(\mathrm{GLM})}=0.070835$ were nearly identical. At $n=1000$, the mean MPLOCO estimate was 0.069198, corresponding to biases of $-0.001769$ and $-0.001637$ relative to the two targets. The mean VIMP estimate was 0.071988, with biases of 0.001021 and 0.001153, respectively. Coverage relative to the intrinsic target was 0.920 for MPLOCO and 0.880 for VIMP; coverage relative to the fixed fitted-function reference was 0.915 and 0.875. Thus, both procedures recovered the same qualitative magnitude of importance when the learner matched the conditional mean model.

The LM results illustrate the distinction between intrinsic importance and importance for a fitted prediction rule. The LM fitted-function reference was $\theta_{0,2}^{(\mathrm{LM})}=0.057063$, substantially below the intrinsic target $\theta_{0,1}=0.070967$. The mean MPLOCO estimate, 0.058780, was close to the learner-specific reference, with a bias of 0.001717 and empirical coverage of 0.865 relative to this reference. In contrast, its bias relative to the intrinsic target was $-0.012187$, and the corresponding empirical coverage was 0.585. This pattern is consistent with the interpretation of MPLOCO as measuring how the fitted learner uses $X_1$: when the learner does not approximate the oracle conditional mean well, its fitted-function importance can differ materially from the population-level intrinsic importance. The mean VIMP estimate was 0.061911 and lay between the two reference values. It remained negatively biased relative to the intrinsic target, with a bias of $-0.009056$, although its wider intervals yielded empirical coverage of 0.900. Relative to the fitted-function reference, its bias and coverage were 0.004849 and 0.910, respectively.

The random-forest results displayed a stronger finite-sample discrepancy. For MPLOCO, the mean estimates were 0.054043, 0.049595, and 0.036273 under the strong, middle, and weak configurations, respectively. All three estimates were below both the intrinsic target and their fitted-function reference values. The weak random forest showed the largest intrinsic bias ($-0.034695$) and the largest MSE$_1$ ($1.224\times10^{-3}$). Coverage was also poor: none of the MPLOCO random-forest configurations exceeded 0.080 relative to the intrinsic target or 0.110 relative to the fitted-function reference. The VIMP random-forest estimate was closer to the intrinsic target (0.059216), but its empirical coverage of 0.575 remained well below the nominal level.

Overall, the results at $n=1000$ support the qualitative distinction developed in Section~2. When a well-specified learner made the intrinsic and fitted-function targets nearly coincide, MPLOCO and VIMP produced estimates on a similar scale. When the learner was misspecified, MPLOCO could instead be close to its learner-specific target while remaining far from the intrinsic target. 

For the logistic regression and random forest learners, the predictions were bounded between zero and one, so squared-error loss was bounded and Lipschitz over the prediction range. 

\subsection{Investigating the importance of groups of features}

The second numerical experiment evaluated the grouped \mploco{} construction introduced in Section~2.5 in a controlled high-dimensional setting motivated by the data analysis in Section~4. The goal was to examine whether grouped \mploco{} could consistently identify important feature groups when groups overlap and the same predictive features may be represented in multiple group definitions. We compared its Monte Carlo behavior with that of grouped VIMP.

For each of 500 Monte Carlo replications, we generated $n=2000$ observations with $p=800$ features. The features were generated independently as
\[X_{ij}\overset{\mathrm{iid}}{\sim}N(0,1),\qquad i=1,\ldots,n,\quad j=1,\ldots,p.\]
The features were organized into $G=14$ prespecified and potentially overlapping groups. Groups 1, 2, and 3 shared a common block of 25 features, and each pair of groups shared an additional pair-specific block. Each of the three groups also contained features unique to that group. Several of the remaining groups overlapped with other noise groups, reflecting the overlapping structure of the biological annotations considered in Section~4.

The outcome was generated according to
\[Y_i\mid X_i\sim\operatorname{Bernoulli}\left[\expit\{\beta_0+X_i^\top\beta\}\right], \qquad\beta_0=\logit(0.25).\]
The 120 features belonging to the four shared blocks associated with groups 1, 2, and 3 were assigned coefficient $\beta_j=0.3$, while all remaining coefficients were zero. Thus, groups 1, 2, and 3 contained the predictive features, whereas groups 4 to 14 contained no predictive features.

Both procedures evaluated the same target groups using differences in mean squared error. For each target group, all features belonging to that group were made unavailable for prediction, as described in Section~\ref{sec:group_vim}. Both procedures used lasso-regularized logistic regression as the prediction algorithm. For grouped \mploco{}, each ensemble contained $B=500$ minipatches, with $n_{\mathrm{ratio}}=0.95$ and $g_{\mathrm{keep}}=5$. Both procedures used pointwise 95\% confidence intervals without multiplicity adjustment.

For each feature group and procedure, we calculated the Monte Carlo mean of the importance estimates, the Monte Carlo standard deviation, and the percentage of confidence intervals lying entirely above zero. Table~\ref{tab:group-mc-all} summarizes the results across the 500 replications.

\begin{table}[!htbp]
\centering
\caption{Monte Carlo summaries of grouped variable-importance estimates
across 500 replications.}
\label{tab:group-mc-all}
\small
\begin{tabular}{lcccc}
\toprule
& \multicolumn{2}{c}{Grouped \mploco{}}
& \multicolumn{2}{c}{Grouped VIMP} \\
\cmidrule(lr){2-3}\cmidrule(lr){4-5}
Group
& Mean (MC SD)
& CI $>0$ (\%)
& Mean (MC SD)
& CI $>0$ (\%) \\
\midrule
\textbf{G1}  & 0.0801 (0.0043)  & 100.0 & 0.0536 (0.0079) & 100.0 \\
\textbf{G2}  & 0.0801 (0.0043)  & 100.0 & 0.0537 (0.0082) & 100.0 \\
\textbf{G3}  & 0.0801 (0.0043)  & 100.0 & 0.0522 (0.0080) & 100.0 \\
\addlinespace
G4  & $-0.0050$ (0.0018) & 0.0 & 0.0016 (0.0034) & 0.8 \\
G5  & $-0.0029$ (0.0010) & 0.2 & 0.0019 (0.0036) & 0.8 \\
G6  & $-0.0046$ (0.0017) & 0.0 & 0.0021 (0.0038) & 0.6 \\
G7  & $-0.0026$ (0.0011) & 0.0 & 0.0022 (0.0037) & 1.0 \\
G8  & $-0.0024$ (0.0011) & 0.0 & 0.0025 (0.0040) & 0.8 \\
G9  & $-0.0041$ (0.0017) & 0.0 & 0.0025 (0.0041) & 1.0 \\
G10 & $-0.0043$ (0.0016) & 0.0 & 0.0026 (0.0041) & 1.0 \\
G11 & $-0.0043$ (0.0016) & 0.0 & 0.0027 (0.0042) & 1.6 \\
G12 & $-0.0050$ (0.0018) & 0.0 & 0.0027 (0.0042) & 1.2 \\
G13 & $-0.0046$ (0.0017) & 0.0 & 0.0028 (0.0042) & 1.0 \\
G14 & $-0.0041$ (0.0017) & 0.0 & 0.0029 (0.0044) & 1.2 \\
\bottomrule
\end{tabular}

\vspace{0.5em}
\begin{minipage}{0.96\linewidth}
\footnotesize
\textit{Note.} Mean (MC SD) denotes the Monte Carlo mean and Monte Carlo standard deviation of the group-importance estimates. The 95\% CI $>0$ column gives the percentage of the 500 pointwise 95\% Wald confidence intervals whose lower endpoint exceeded zero. Groups G1--G3, shown in bold, contained the predictive features. No multiplicity adjustment was applied.
\end{minipage}
\end{table}

Both procedures consistently identified the three groups containing predictive features. For grouped \mploco{}, groups 1, 2, and 3 each had a Monte Carlo mean estimate of $0.0801$ and a Monte Carlo standard deviation of $0.0043$. For grouped VIMP, their mean estimates ranged from $0.0522$ to $0.0537$, with Monte Carlo standard deviations ranging from $0.0079$ to $0.0082$. For both procedures, the confidence intervals for each of the three signal-containing groups were entirely above zero in all 500 replications. Equivalently, all three signal-group confidence intervals were simultaneously above zero in every replication.

The identical grouped \mploco{} results for groups 1, 2, and 3 are a consequence of the overlapping group construction. Under the grouped leave-out rule, a minipatch was eligible for the reduced prediction only when it contained none of the features in the target group. Because each pair among groups 1, 2, and 3 shared predictive features, the three targets induced the same collection of eligible leave-group-out minipatches. Grouped \mploco{} therefore identified the shared signal-containing group structure, but did not distinguish the individual contributions of these three highly overlapping groups.

The remaining groups showed substantially smaller estimates. For grouped \mploco{}, the Monte Carlo mean estimates for groups 4--14 ranged from $-0.0050$ to $-0.0024$, with Monte Carlo standard deviations between $0.0010$ and $0.0018$. Their confidence intervals were almost never entirely above zero: the corresponding percentages ranged from $0\%$ to $0.2\%$. These small negative estimates do not imply negative scientific importance. Rather, for the fitted minipatch ensembles, excluding these noise groups did not increase out-of-minipatch prediction error on average.

For grouped VIMP, the mean estimates for groups 4--14 were close to zero, ranging from $0.0016$ to $0.0029$, with Monte Carlo standard deviations between $0.0034$ and $0.0044$. The percentage of confidence intervals lying entirely above zero ranged from $0.6\%$ to $1.6\%$. Thus, grouped VIMP also provided little evidence of positive importance for the groups containing no predictive features.

Overall, in this illustrative simulated dataset, grouped \mploco{} successfully identified the same three signal-containing groups as grouped VIMP. The two procedures produced different numerical estimates and interval widths, but these differences did not alter the principal ranking of the groups. This experiment therefore provides an initial controlled illustration of the grouped \mploco{} procedure proposed in Section~2.5.

\section{Importance of HIV-1 genetic features for antibody neutralization}

We analyzed a VRC01 neutralization sensitivity dataset to compare intrinsic VIMP and extrinsic \mploco{} in a high-dimensional biological prediction problem. This dataset was studied by \citet{magaret2019prediction} and later used as a real-data illustration of intrinsic variable importance by \citet{williamson2023general}. Our goal was not to conduct a new biological discovery study, but rather to compare the two notions of variable importance in a scientifically meaningful application. The dataset is well suited for this comparison because it contains many correlated amino acid sequence features, several biologically motivated and overlapping feature groups, and a binary outcome with direct scientific interpretation.

VRC01 is a broadly neutralizing monoclonal antibody targeting the HIV-1 envelope protein and was developed as part of a broader pathway toward antibody-mediated HIV-1 prevention \citep{gilbert2017basis,corey2021two}. A given HIV-1 virus may be sensitive or resistant to neutralization by VRC01, and this sensitivity depends in part on amino acid sequence features of the HIV-1 envelope protein \citep{magaret2019prediction}. We analyzed HIV-1 envelope amino acid sequence features from 611 publicly available HIV-1 Env pseudoviruses derived from blood samples of HIV-1-infected individuals \citep{magaret2019prediction}. For each pseudovirus, the available data included amino acid sequence features of the HIV-1 envelope protein, geographic information, and in vitro neutralization measurements. Because the number of residue-level sequence features was large relative to the sample size, this dataset provided a challenging setting for variable importance inference.

The outcome of interest was a binary indicator of VRC01 neutralization sensitivity, defined using the geometric mean VRC01 50\% inhibitory concentration (IC50). Following previous analyses, viruses with geometric mean IC50 less than 1 were classified as sensitive, while viruses with larger IC50 values were classified as resistant \citep{magaret2019prediction,williamson2023general}. Thus, the prediction task was to classify whether an HIV-1 Env pseudovirus was sensitive to VRC01 neutralization using amino acid sequence and related biological features.

Following previous analyses of this dataset, we considered approximately 800 individual amino acid sequence features, together with 14 biologically motivated feature groups. These groups summarized regions or mechanisms potentially related to VRC01 neutralization, including the VRC01 binding footprint, CD4 binding sites, sites with sufficient exposed surface area, glycosylation-related sites, subtype-related features, viral geometry, cysteine counts, steric bulk measures, and geographic covariates \citep{magaret2019prediction,williamson2023general}. Because of the large number of individual residue-level features, our primary analysis focused on grouped variable importance for all prespecified biological feature groups. We then performed a secondary individual-feature analysis for variables within the VRC01 binding footprint group. This group was selected because mutations at amino acid positions in the VRC01 binding footprint are expected to affect VRC01 neutralization potential \citep{magaret2019prediction,zhou2010structural}.

We estimated grouped variable importance under both the intrinsic VIMP framework and the extrinsic \mploco{} framework, using lasso-regularized logistic regression as the prediction algorithm. As in the numerical experiments, variable importance was defined using differences in mean squared error, placing the two procedures on the same loss scale. For \mploco{}, we used $B=2000$ minipatches, sampled 95\% of the observations in each minipatch, and sampled five of the 14 feature groups. For VIMP, we used two-fold sample splitting to estimate the full and reduced prediction risks.

To use a common nonnegative reporting convention across the two procedures, negative finite-sample \mploco{} point estimates were truncated at zero. When truncation occurred, we retained the estimate standard error and constructed a Wald-type confidence interval centered at zero. Consequently, a point estimate displayed at zero may have a confidence interval extending below zero. Both procedures used pointwise 90\% confidence intervals without multiplicity adjustment.

\begin{figure}[htbp]
\centering
\includegraphics[width=\textwidth]{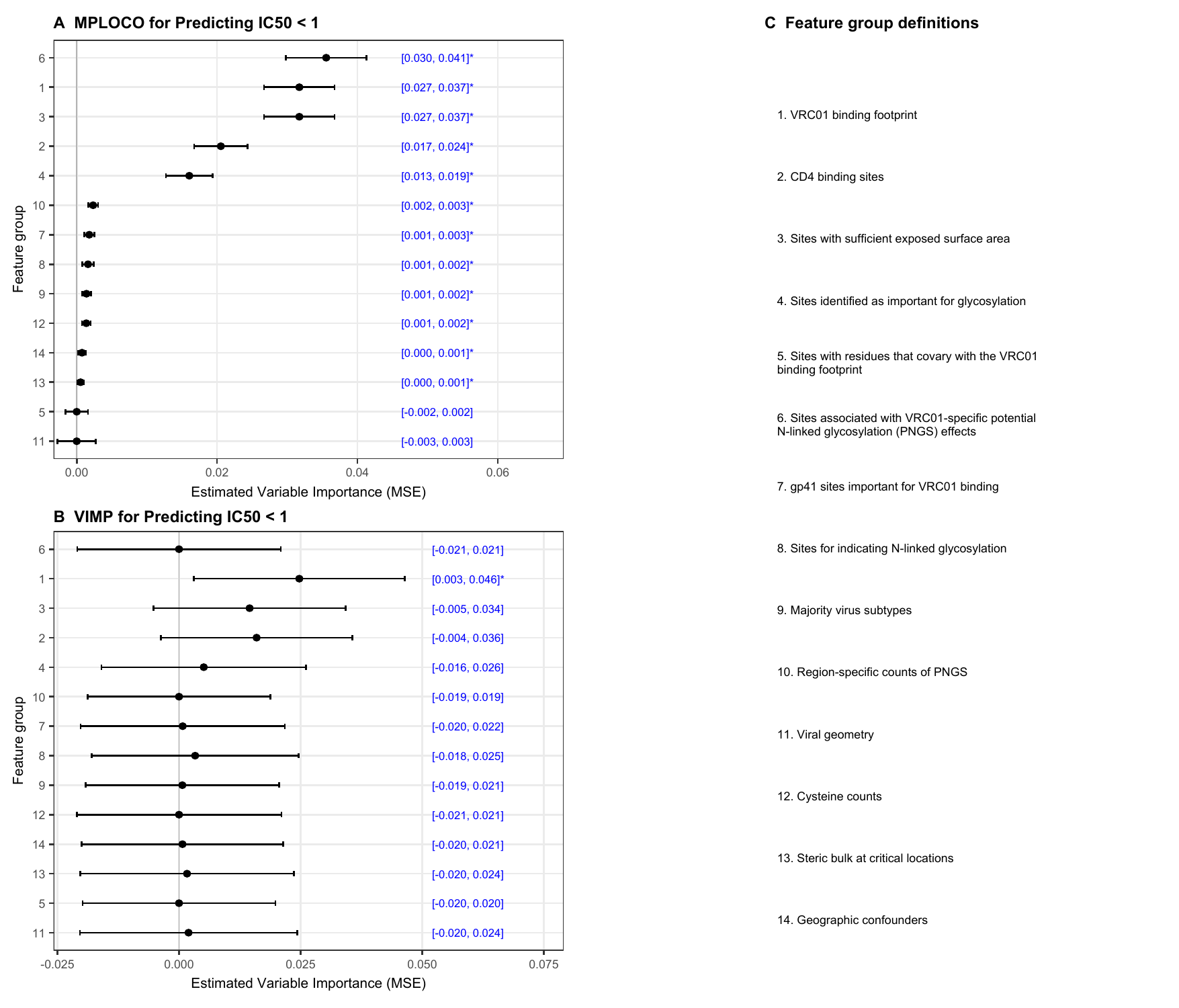}
\caption{
Grouped variable importance estimates for predicting VRC01 neutralization sensitivity, defined as IC50 $<1$. Panel A shows estimates from the lasso-based \mploco{} procedure, Panel B shows estimates from VIMP, and Panel C lists the biological feature-group definitions. Variable importance is measured using differences in mean squared error. Pointwise 90\% confidence intervals are shown, and intervals excluding zero are marked with an asterisk. }
\label{fig:vrc01_grouped}
\end{figure}

Figure~\ref{fig:vrc01_grouped} compares the grouped variable importance estimates from the two procedures. Groups 1 and 2 were expected a priori to be important. Group 1 corresponds to the VRC01 binding footprint, while group 2 corresponds to CD4 binding sites; both are directly related to VRC01 recognition of HIV-1 Env \citep{magaret2019prediction,zhou2010structural}. Both procedures ranked these groups among the most important feature groups, providing evidence of a biologically meaningful predictive signal related to direct antibody binding and viral entry mechanisms.

A notable additional result was the high ranking of group 3, corresponding to sites with sufficient exposed surface area. This group was ranked highly by both procedures, even though it was not as direct an a priori target as the VRC01 binding footprint or CD4 binding sites. This result is biologically plausible because surface-accessible residues may affect whether antibody-relevant regions are accessible for neutralization \citep{magaret2019prediction,zhou2010structural}. Thus, the results suggest that predictive information for VRC01 sensitivity is not limited to predefined binding regions, but may also include broader structural accessibility features.

The \mploco{} analysis also assigned a relatively large importance estimate to group 6, corresponding to sites associated with VRC01-specific potential N-linked glycosylation effects. We interpret this difference cautiously. The sample size was modest relative to the number of sequence features, and the biological feature groups were correlated and overlapping. The large \mploco{} estimate for group 6 therefore suggests that glycosylation-related features were useful to the fitted lasso-based minipatch ensemble, possibly because they captured predictive information that overlapped with or complemented information represented by other biological groups.

Comparing Panels A and B of Figure~\ref{fig:vrc01_grouped} also shows differences in the numerical estimates and interval widths. The \mploco{} intervals were generally narrower, and more groups had intervals excluding zero, whereas the VIMP intervals were wider. Under the alignment conditions developed in Section~2, the fitted-learner importance target can approach the intrinsic population-level VIMP target. However, in this modest-sample, high-dimensional application, these conditions may not hold closely in finite samples. The observed differences may therefore reflect both the construction of the fitted learners and the different inferential procedures. They should not be interpreted as showing that one method is uniformly preferable to the other.

Several feature groups showed little evidence of importance under both analyses. For example, group 11, corresponding to viral geometry, had an estimate near zero under \mploco{} and a confidence interval centered near zero under VIMP. Similar patterns were observed for several lower-ranked groups, including cysteine counts, steric bulk measures, and geographic covariates. These results suggest that these groups provided relatively little additional predictive information for the binary neutralization outcome after accounting for the more highly ranked binding, exposure, and glycosylation-related features.

\begin{figure}[H]
\centering
\includegraphics[width=\textwidth]{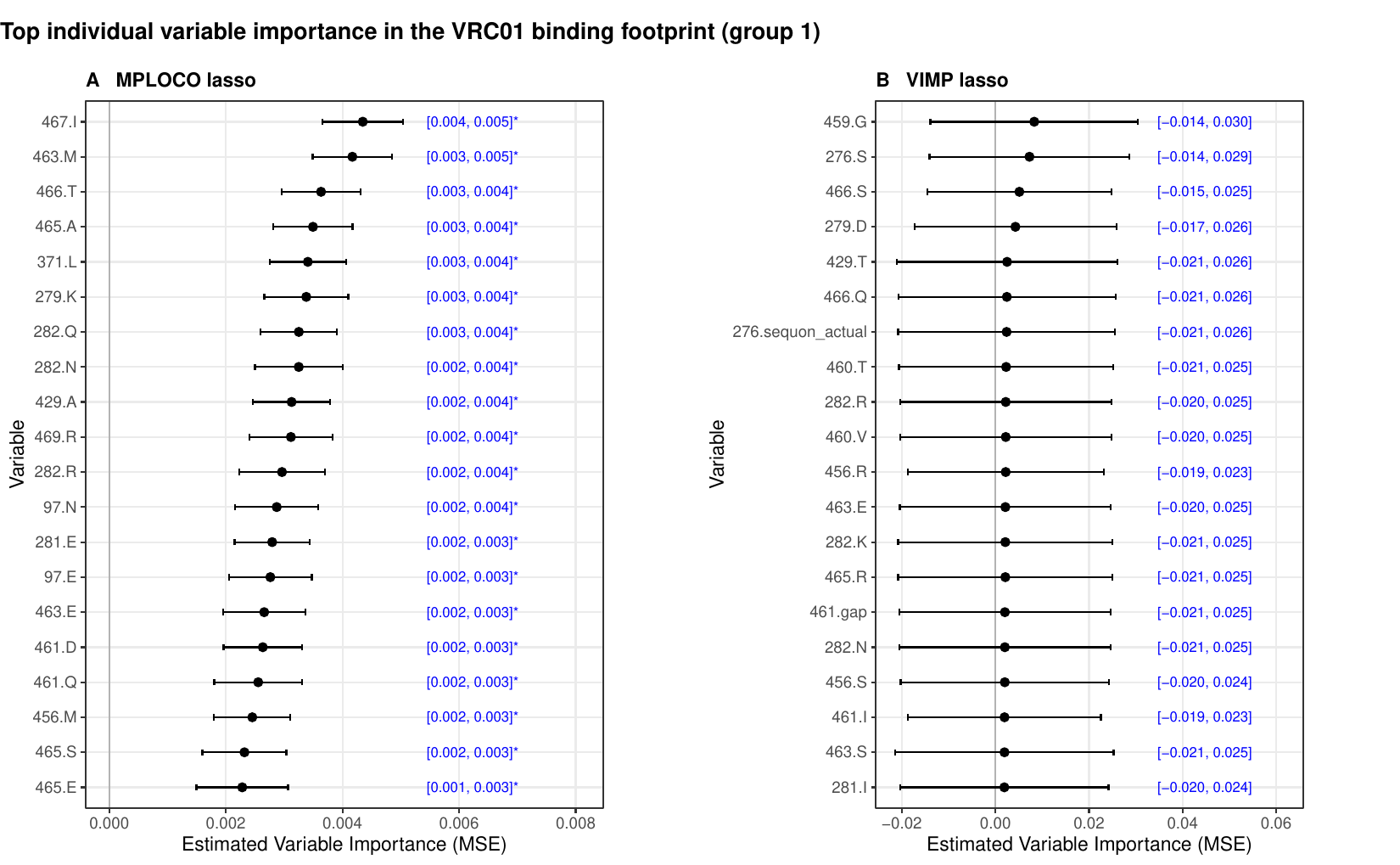}
\caption{
Top individual variable importance estimates within the VRC01 binding footprint group (group 1).
Panel A shows estimates from the MPLOCO procedure, and Panel B shows estimates from the VIMP procedure.
The top 20 variables were selected separately for the two procedures.
Variable labels are written as amino acid position followed by residue identity; for example, 467.I denotes residue I at HXB2 position 467.
Variable importance is measured using differences in mean squared error.
}
\label{fig:vrc01_individual}
\end{figure}

To further examine predictive signal within an important biologically defined group, Figure~\ref{fig:vrc01_individual} compares the top individual variables within group 1, the VRC01 binding footprint, under \mploco{} and VIMP. The top 20 variables were selected separately for the two procedures. Several residue-level features appeared in both panels or in closely related forms, including variables involving positions 463, 465, and 466. This overlap suggests that the procedures identified a partially shared subset of VRC01 binding-footprint residues as predictive of neutralization sensitivity.

At the individual-feature level, the two approaches diverged more than they did at the group level. Under lasso-based \mploco{}, many individual variables had small positive estimates with narrow intervals, whereas under VIMP the corresponding intervals were wider and often included zero. This pattern is consistent with predictive information being distributed across multiple correlated residue-level variables. Grouped variable importance therefore provided a more stable and interpretable summary of the biological signal than attempting to assign importance to each individual residue separately.

Overall, the analysis supports two main conclusions. First, at the group level, both \mploco{} and VIMP ranked feature groups with clear biological relevance to VRC01 neutralization highly, including the VRC01 binding footprint, CD4 binding sites, and exposed-surface-area features. At the individual level, the procedures identified partially overlapping sets of residue-level variables within the VRC01 binding footprint that are plausibly related to neutralization sensitivity. Second, the two procedures provided complementary information. Differences in the top-ranked groups and individual features, including the larger \mploco{} estimate for the VRC01-specific glycosylation group, illustrate how the fitted learner and the inferential construction can affect variable importance conclusions in a modest-sample, high-dimensional setting.

\section{Conclusions}

In this paper, we clarified the relationship between intrinsic population-level VIMP and extrinsic fitted-learner importance estimated by \mploco{}. Although both procedures compare predictive performance with and without a feature or feature group, they begin from different inferential objects. VIMP defines importance through oracle prediction functions under the population distribution, whereas \mploco{} evaluates the contribution of a feature to a fitted minipatch ensemble. Our theoretical results show that these two variable importance targets, and the asymptotic variance of estimators of these targets, align under conditions on the rate at which the fitted full and reduced prediction functions approach their oracle counterparts. When these conditions are not satisfied, differences in estimates or uncertainty should be interpreted as reflecting learner-specific and finite-sample behavior rather than as a contradiction between the procedures. In all cases, the two procedures provide complementary information about the importance of variables in population and the fitted model.

The numerical and data examples illustrated how this distinction affects interpretation in practice. In the grouped simulated example, both procedures ranked the three groups containing predictive signal above the remaining groups, providing an initial controlled illustration of grouped \mploco{} with overlapping feature definitions. In the VRC01 analysis, both procedures ranked biologically relevant groups highly, including the VRC01 binding footprint, CD4 binding sites, and exposed-surface-area features, while differences in other group rankings and interval widths highlighted the roles of the fitted learner and inferential construction. These findings should be interpreted cautiously because our analyses focused on mean-squared-error importance, a limited collection of prediction algorithms and data-generating settings, and a modest-sample, high-dimensional application with correlated and overlapping features. Overall, population-level and fitted-learner importance answer related but distinct questions, and clarifying when they align is essential for interpreting variable-importance procedures in scientific applications.


\bibliographystyle{chicago}
\bibliography{references}

\end{document}


\maketitle

\doublespacing

\section{Proofs of theorems}
\label{sec:supplement}

\subsection{Estimator-specific inference conditions}
\label{sec:supp_inference_conditions}

Condition (C3) in the main manuscript collects the separate assumptions needed for valid inference on the intrinsic VIMP target and on the fitted \mploco{} target. We state those assumptions here using the notation of the main manuscript.

\paragraph{Conditions for VIMP inference.}

For the cross-fitted VIMP estimator, the conditions of \citet{williamson2023general} are applied to both the full and reduced prediction problems. Let $\widehat f_n^{(-k)}$ and $\widehat f_{n,-s}^{(-k)}$ denote the learners trained without validation fold $k$. The relevant conditions are as follows.

\begin{enumerate}
\item[(V1)] \textbf{Local quadratic optimality.}
In neighborhoods of the full and reduced oracle functions, there is a constant $C<\infty$ such that \[\left|R_0(f)-R_0(f_0)\right|\leq C\|f-f_0\|_{L_2(P_0)}^2\] and \[\left|R_0(f_{-s})-R_0(f_{0,-s})\right|\leq C\|f_{-s}-f_{0,-s}\|_{L_2(P_0)}^2.\]

\item[(V2)] \textbf{Differentiability of the risk functional.}
For fixed $f$, the map $P\mapsto R_P(f)$ is differentiable at $P_0$ along regular submodels, and its derivative is continuous as $f$ approaches the corresponding oracle function. In particular, writing $\dot R_0(f;h)$ for the derivative in direction $h$, the evaluations \[\dot R_0(f;\delta_z-P_0)\] must be well defined and locally continuous in $f$ in the norm used for the influence-function expansion. The analogous condition is required for the reduced risk.

\item[(V3)] \textbf{Fold-specific oracle rates.}
Condition (C1) holds for each training fold:
\[\|\widehat f_n^{(-k)}-f_0\|_{L_2(P_0)}=o_P(n^{-1/4})\] and \[\|\widehat f_{n,-s}^{(-k)}-f_{0,-s}\|_{L_2(P_0)}=o_P(n^{-1/4})\] for every fixed $k$.

\item[(V4)] \textbf{Weak consistency of the influence-function contributions.}
For every fixed validation fold $k$, \[\begin{aligned}\int\Big[&\dot R_0\{\widehat f_n^{(-k)};\delta_z-P_0\}\\&-\dot R_0(f_0;\delta_z-P_0)\Big]^2dP_0(z)=o_P(1),\end{aligned}\] with the corresponding condition also holding for the reduced learner.

\item[(V5)] \textbf{Sampling and variance conditions.} The observations are i.i.d., the number of cross-fitting folds is fixed with fold proportions bounded away from zero, Condition (C4) holds, and the empirical second moment of the estimated influence-function contributions consistently estimates $\sigma_1^2$.
\end{enumerate}

Under cross-fitting, these conditions replace the Donsker-class restriction that would otherwise be imposed on the fitted learners. They yield the VIMP limit and variance consistency stated in Condition (C3). Conditions concerning smooth variation of the population optimizer and continuity of the derivative are additionally needed to identify the resulting influence function as the nonparametric efficient influence function, but that efficiency designation is not needed for the alignment theorem.

For squared-error loss, \[\ell\{y,f(x)\}=\{y-f(x)\}^2,\] Condition (V1) holds with equality because \[R_0(f)-R_0(f_0)=\|f-f_0\|_{L_2(P_0)}^2,\] and similarly for the reduced oracle. Moreover, for a fixed $f$,\[\dot R_0(f;\delta_z-P_0)=\ell\{y,f(x)\}-R_0(f).\] Consequently, bounded outcomes and predictions, or the fourth-moment conditions described in Section~\ref{sec:supp_c2_conditions}, give the required local continuity and weak consistency. Mean absolute error is not used in the main theorem because its risk is generally nonsmooth at a zero residual and need not satisfy the local quadratic expansion in (V1) without additional assumptions.

\paragraph{Conditions for \mploco{} fitted-target inference.}

Let $r_n=|I_b|$ and $q_n=|F_b|$ denote the observation and feature minipatch sizes, respectively, and let $B_n$ denote the number of sampled minipatches. Let $D<\infty$ bound the difference between predictions obtained from any two admissible minipatches, and let $L<\infty$ be the Lipschitz constant of the loss. In the notation of the main manuscript, the source conditions summarized from \citet{gan2022model} are:

\begin{enumerate}
\item[(M1)] \textbf{Independent sampling.}
$Z_1,\ldots,Z_n$ are i.i.d.\ from $P_0$.

\item[(M2)] \textbf{Lipschitz loss.}
For all relevant $y,a,b$, \[\left|\ell(y,a)-\ell(y,b)\right|\leq L|a-b|.\]

\item[(M3)] \textbf{Bounded minipatch prediction differences.}
For any two admissible observation--feature minipatches $(I,F)$ and $(I',F')$ and every relevant input $x$, \[\left|\widehat f_{I,F}(x)-\widehat f_{I',F'}(x)\right|\leq D.\]

\item[(M4)] \textbf{Minipatch-size rates.}
For some constant $\beta\in(0,1)$, \[\frac{r_n}{n}\leq\beta,\qquad\frac{q_n}{p}\leq\beta,\] and \[r_n=o\left(\frac{\sigma_{2,n}\sqrt n}{LD}\right).
\]

\item[(M5)] \textbf{Number of minipatches.}
The number of sampled minipatches satisfies \[B_n\gg\left(\frac{L^2D^2n}{\sigma_{2,n}^2}+\frac{LD\sqrt n}{\sigma_{2,n}}+1\right)\log n.\]

\item[(M6)] \textbf{Uniform integrability.}
The standardized squared observation-level importance quantities \[\left\{\frac{\left[g_n(Z)-E_0\{g_n(Z)\mid\mathcal D_n\}\right]^2}{\sigma_{2,n}^2}\right\}_{n\geq1}\] are uniformly integrable.
\end{enumerate}

These conditions yield the fitted-target central limit theorem and variance consistency invoked in Condition (C3). They are separate from Conditions (C1) and (C2), which connect the fitted target and variance to their oracle counterparts.

\paragraph{Squared-error loss under the \mploco{} conditions.}

Squared-error loss is not globally Lipschitz on an unbounded prediction range. If $|Y|\leq M_Y$ and $|a|,|b|\leq M_f$, however, then \[\begin{aligned}\left|(Y-a)^2-(Y-b)^2\right|&=|a-b||a+b-2Y|\\&\leq2(M_f+M_Y)|a-b|.\end{aligned}\] Thus, bounded outcomes and uniformly bounded or clipped predictions verify (M2), and they also imply (M3) with $D\leq2M_f$. The minipatch-size, ensemble-size, and uniform-integrability requirements remain separate conditions; squared-error loss by itself does not establish them.

\subsection{Sufficient conditions for Condition (C2)}
\label{sec:supp_c2_conditions}

We now make explicit when the loss-contrast convergence in Condition (C2) follows under squared-error loss. Write \[g_n-g_0=A_n-B_n,\] where \[\begin{aligned}A_n={}&\{Y-f_{n,-s}(X_{-s})\}^2-\{Y-f_{0,-s}(X_{-s})\}^2,\\B_n={}&\{Y-f_n(X)\}^2-\{Y-f_0(X)\}^2.\end{aligned}\]

\paragraph{Bounded sufficient condition.}

Suppose that, with probability tending to one, there are constants $M_Y,M_f<\infty$ such that \[ |Y|\leq M_Y\] and all four fitted and oracle predictions are bounded in absolute value by $M_f$. Then \[\begin{aligned}\|B_n\|_{L_2(P_0)}&\leq2(M_f+M_Y)\|f_n-f_0\|_{L_2(P_0)},\\\|A_n\|_{L_2(P_0)}&\leq2(M_f+M_Y)\|f_{n,-s}-f_{0,-s}\|_{L_2(P_0)}.\end{aligned}\] Therefore, by the triangle inequality and Condition (C1),\[\begin{aligned}\|g_n-g_0\|_{L_2(P_0)}\leq{}&2(M_f+M_Y)\Big\{\|f_n-f_0\|_{L_2(P_0)}\\&+\|f_{n,-s}-f_{0,-s}\|_{L_2(P_0)}\Big\}\overset{P}{\longrightarrow} 0.\end{aligned}\] Hence the bounded sufficient condition does hold for squared-error loss.

\paragraph{Moment sufficient condition.}

Boundedness can be replaced by stronger convergence and moment conditions. By H\"older's inequality, \[\|B_n\|_{L_2(P_0)}\leq\|f_n-f_0\|_{L_4(P_0)}\|2Y-f_n-f_0\|_{L_4(P_0)},\] and the analogous inequality holds for $A_n$. It is therefore sufficient that \[\|f_n-f_0\|_{L_4(P_0)}\overset{P}{\longrightarrow} 0,\qquad\|f_{n,-s}-f_{0,-s}\|_{L_4(P_0)}\overset{P}{\longrightarrow}0,\] and that the $L_4(P_0)$ norms of $Y$, the fitted predictions, and the oracle predictions are bounded in probability. These conditions imply Condition (C2). Without boundedness, the $L_2(P_0)$ rates in Condition (C1) alone do not imply the required $L_2(P_0)$ convergence of the squared-loss contrasts.

\subsection{Proof of
Theorem~\ref{thm:vimp_mploco_alignment}}
\label{sec:supp_alignment_proof}

\begin{proof}
Under squared-error loss, the conditional-mean oracle property gives the exact identities \[R_0(f_n)-R_0(f_0)=\|f_n-f_0\|_{L_2(P_0)}^2\] and \[R_0(f_{n,-s})-R_0(f_{0,-s})=\|f_{n,-s}-f_{0,-s}\|_{L_2(P_0)}^2.\] For example, \[\begin{aligned}R_0(f_n)-R_0(f_0)={}&E_0\left[\{Y-f_n(X)\}^2-\{Y-f_0(X)\}^2\mid\mathcal D_n\right]\\={}&\|f_n-f_0\|_{L_2(P_0)}^2\\&+2E_0\left[\{Y-f_0(X)\}\{f_0(X)-f_n(X)\} \mid\mathcal D_n\right].\end{aligned}\] The last expectation is exactly zero because $E_0\{Y-f_0(X)\mid X\}=0$. The reduced identity follows in the same way after conditioning on $X_{-s}$.

Consequently, \[\begin{aligned}\theta_{0,2}(f_n)-\theta_{0,1}={}&\|f_{n,-s}-f_{0,-s}\|_{L_2(P_0)}^2\\&-\|f_n-f_0\|_{L_2(P_0)}^2.\end{aligned}\] Condition (C1) therefore implies \[\begin{aligned}\sqrt n\left|\theta_{0,2}(f_n)-\theta_{0,1}\right|\leq{}&\sqrt n\|f_{n,-s}-f_{0,-s}\|_{L_2(P_0)}^2\\&+\sqrt n\|f_n-f_0\|_{L_2(P_0)}^2\\={}&o_P(1).\end{aligned}\] This proves part (ii), and part (i) follows immediately.

For part (iii), let \[d_n=g_n-g_0.\] Condition (C2) gives \[\|d_n\|_{L_2(P_0)}\overset{P}{\longrightarrow} 0.\] Using the fitted-contrast variance \[\sigma_{2,n}^2=\operatorname{Var}_0\{g_n(Z)\mid\mathcal D_n\}\] that appears in Condition (C3), we obtain \[\begin{aligned}\left|\sigma_{2,n}^2-\sigma_1^2\right|&\leq\operatorname{Var}_0\{d_n(Z)\mid\mathcal D_n\}\\&\quad+2\left|\operatorname{Cov}_0\{g_0(Z),d_n(Z)\mid\mathcal D_n\}\right|\\&\leq\|d_n\|_{L_2(P_0)}^2+2\sigma_1\|d_n\|_{L_2(P_0)}\\&\overset{P}{\longrightarrow} 0.\end{aligned}\] Condition (C3) also gives \[\frac{\widehat\sigma_{2,n}^2}{\sigma_{2,n}^2} \overset{P}{\longrightarrow} 1.\] Together with Condition (C4) and $\sigma_{2,n}^2\overset{P}{\longrightarrow}\sigma_1^2$, this implies \[\widehat\sigma_{2,n}^2\overset{P}{\longrightarrow}\sigma_1^2.\]

Finally,\[\begin{aligned}\sqrt n\left(\widehat\theta_{n,2}-\theta_{0,1}\right)={}&\sqrt n\left\{\widehat\theta_{n,2}-\theta_{0,2}(f_n)\right\}\\&+\sqrt n\left\{\theta_{0,2}(f_n)-\theta_{0,1}\right\}.\end{aligned}\] By Condition (C3) and $\sigma_{2,n}^2\overset{P}{\longrightarrow}\sigma_1^2>0$, the first term converges in distribution to $N(0,\sigma_1^2)$. The second term is $o_P(1)$ by part (ii). Slutsky's theorem proves part (iv). 
\end{proof}

\subsection{Extension beyond squared-error loss}
\label{sec:supp_general_loss}

\begin{theorem}[General-loss alignment]
\label{thm:general_loss_alignment}
Suppose Conditions (C2)--(C4) hold and \[\sqrt n\left[\left\{R_0(f_{n,-s})-R_0(f_{0,-s})\right\}-\left\{R_0(f_n)-R_0(f_0)\right\}\right]=o_P(1).\] Then \[\sqrt n\left\{\theta_{0,2}(f_n)-\theta_{0,1}\right\}=o_P(1),\] \[\sigma_{2,n}^2\overset{P}{\longrightarrow}\sigma_1^2,\qquad\widehat\sigma_{2,n}^2\overset{P}{\longrightarrow}\sigma_1^2,\] and \[\sqrt n\left(\widehat\theta_{n,2}-\theta_{0,1}\right)\overset{d}{\longrightarrow} N(0,\sigma_1^2).\]
\end{theorem}

\begin{proof}
The assumed excess-risk condition is exactly the root-$n$ target alignment \[\sqrt n\left\{\theta_{0,2}(f_n)-\theta_{0,1}\right\}=o_P(1).\] The variance argument follows from Condition (C2) exactly as in the proof of part (iii) of Theorem~\ref{thm:vimp_mploco_alignment}. Condition (C3), the root-$n$ target alignment, and Slutsky's theorem then give the stated limiting distribution.
\end{proof}

\bibliographystyle{chicago}
\bibliography{references}